\documentclass[
    ,final            
  ]
  {aipproc}
\usepackage{epsfig}

\layoutstyle{6x9}

\begin{document}

\title{Excitation of trapped oscillations in discs around black holes}


\author{B\'arbara T. Ferreira}{
  address={Department of Applied Mathematics and Theoretical Physics, University of Cambridge, Centre for Mathematical Sciences, Wilberforce Road, Cambridge CB3 0WA, UK}
}

\author{Gordon I. Ogilvie}{
  address={Department of Applied Mathematics and Theoretical Physics, University of Cambridge, Centre for Mathematical Sciences, Wilberforce Road, Cambridge CB3 0WA, UK}
}

\begin{abstract}
High-frequency quasi-periodic oscillations detected in the light curves of black hole candidates can, according to one model, be identified with hydrodynamic oscillations of the accretion disc. We describe a non-linear coupling mechanism, suggested by Kato, through which inertial waves trapped in the inner regions of accretion discs around black holes are excited. Global warping and/or eccentricity of the disc have a fundamental role in this coupling: they combine with trapped modes, generating negative energy waves, that are damped as they approach the inner edge of the disc or their corotation resonance. As a result of this damping, inertial oscillations are amplified. We calculate the resulting eigenfunctions and their growth rates.
\end{abstract}

\maketitle

\section{1. Introduction}

The idea of using wave coupling as an excitation mechanism for oscillations in deformed discs goes back to Goodman \citep{goodman1993}, who studied the excitation of waves in tidally deformed discs via the parametric instability. Papaloizou \& Terquem \citep{papterquem1995} also mention the possibility of parametric generation of inertial waves in warped discs. More recently, an analytical study of the excitation of waves in warped and eccentric discs around non-rotating black holes was carried out by Kato \citep{katowarp2004,kato2007}. His estimates for the growth rates of several modes in the disc are interesting, but many aspects of his calculations are uncertain, and the origin and dynamic properties of the global warping or eccentricity are not discussed.

In the work reported here, Kato's ideas are developed and generalised for discs around rotating black holes, and detailed numerical calculations of trapped inertial modes and their growth rates are performed. A dynamical treatment of global deformations is included; their origin is discussed in Ogilvie \& Ferreira (this volume).

In $\S 2$ we discuss the trapping of inertial waves, in $\S 3$ we describe the coupling mechanism through which these waves are excited, in $\S 4$ we discuss the variation of the inertial mode growth rate with several parameters, and we present our conclusions in $\S 5$.

\section{2. Trapped oscillations}

Trapping of oscillations \citep{katofukue1980} is the result of the non-monotonic variation of the epicyclic frequency $\kappa$ in relativistic discs. As the radius decreases, $\kappa$ increases, reaching a maximum, and then decreases rapidly, going to zero at the marginally stable orbit (regarded as the inner edge of the disc). Analysis of the hydrodynamic equations in a simple disc model \citep{lubowpringle1993} shows that inertial waves can be trapped below this maximum \citep{okazakietal1987} \citep[see also][]{rkato2001,nowaklehrchapter}.

Here we consider an isothermal disc with a ratio of specific heats $\gamma=1$. We ignore viscosity and magnetic fields, and relativistic effects are included by using relativistic expressions for the angular velocity $\Omega$, vertical frequency $\Omega_z$, and $\kappa$ \citep{kato1990}. Perturbations about the equilibrium disc can be described by

\begin{equation}
(u'_{r},u'_{\phi},h')=\textrm{Re}\left\{\left[u_{r}(r),u_{\phi}(r),h(r)\right]\textrm{He}_n\left(\frac{z}{H}\right)\exp(\textrm{i}m\phi-\textrm{i}\omega t)\right\},
\end{equation}
\begin{equation}
u'_{z}=\textrm{Re}\left\{u_{z}(r)\textrm{He}_{n-1}\left(\frac{z}{H}\right)\exp(\textrm{i}m\phi-\textrm{i}\omega t)\right\},
\end{equation}
where $(u'_{r},u'_{\phi},u_z')$ are the components of the velocity perturbation, $h'$ is the enthalpy perturbation, $m$ and $n=0,1,2,...$ are the azimuthal and vertical wave numbers, respectively, $\omega$ is the oscillation frequency, and $H=c_s/\Omega_z$ is the vertical scale-height of the disc, with $c_s$ being the constant sound speed; $(r,\phi,z)$ are cylindrical polar coordinates. The radial structure of linear oscillations is then described approximately by the following set of ordinary differential equations in $r$

\begin{equation}
-\mathrm{i}\hat{\omega}u_r - 2\Omega u_\phi=-\frac{\mathrm{d}h}{\mathrm{d}r},
\label{free1}
\end{equation}
\begin{equation}
-\mathrm{i}\hat{\omega}u_\phi +\frac{\kappa^2}{2\Omega}u_r =-\frac{\mathrm{i}mh}{r},
\end{equation}
\begin{equation}
-\mathrm{i}\hat{\omega}u_z =-n\frac{h}{H},
\end{equation}
\begin{equation}
-\mathrm{i}\hat{\omega}h-\Omega_z^2Hu_z=-c_\mathrm{s}^2\left[\frac{1}{r}\frac{\mathrm{d} (ru_r)}{\mathrm{d} r}+\frac{\mathrm{i}mu_\phi}{r}\right],
\label{free2}
\end{equation}
where $\hat{\omega}=\omega-m\Omega$ is the Doppler-shifted wave frequency, which is zero at the corotation resonance.

The dispersion relation for waves with local radial wavenumber $k$, can be obtained from these equations by assuming the radial wavelength of oscillations to be much smaller than both the azimuthal wavelength and characteristic radial scale in the equilibrium state; it reads

\begin{equation}
k^2=\frac{(\hat{\omega}^2-\kappa^2)(\hat{\omega}^2-n\Omega_z^2)}{\hat{\omega}^2c_\mathrm{s}^2}.
\label{disprelation}
\end{equation}

From this relation we can determine the different types of waves existing in a disc, as well as their propagation regions. If $n\neq0$, acoustic (p) modes can propagate where $\hat{\omega}^2>\textrm{max}(\kappa^2,n\Omega_z^2)=n\Omega_z^2$, and inertial (r) modes with
$\hat{\omega}^2<\textrm{min}(\kappa^2,n\Omega_z^2)=\kappa^2$ can exist. If $n=0$, waves can
propagate where $\hat{\omega}^2>\kappa^2$; this is the inertial-acoustic mode.

Here we focus on the simplest possible inertial mode ($m=0$, $n=1$ and simple radial structure), with frequency $\omega\approx\textrm{max}(\kappa)$, which is confined within a small region below the maximum of the epicyclic frequency, making it more likely to survive in the presence of turbulent viscosity. An axisymmetric mode is likely to be the most relevant observationally, as it results in no cancellations in the net luminosity variations of the disc. The fact that the frequency of this mode can be identified with $\textrm{max}(\kappa)$ is important if one is interested in studying the properties of the black hole, since the maximum of the epicyclic frequency depends only on the mass $M$ and spin $a$ of the compact object \citep{nowaketal1997}. 

\begin{figure}	
  \includegraphics[width=2.86 in]{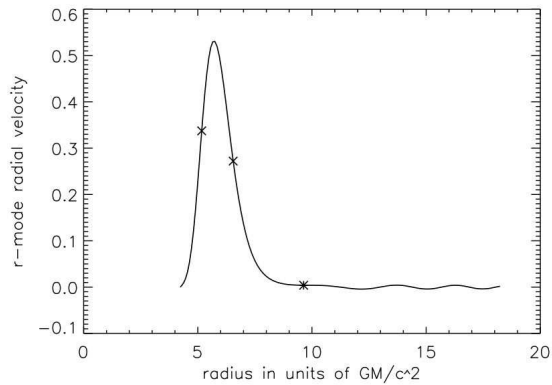}
  \caption{Variation with radius of the real part of the radial velocity of the simplest possible r mode $(l,m,n)=(0,0,1)$, for $c_s/c=0.01, a=0.5$. The maximum is achieved at the radius where $\kappa$ is also maximised. The crosses and asterisks indicate the Lindblad ($\omega^2=\kappa^2$) and vertical ($\omega^2=\Omega_z^2$) resonances, respectively.}
\label{freer}
\end{figure}

To find the radial structure of axisymmetric modes with $n=1$, we solve equations (\ref{free1})-(\ref{free2}) numerically \citep{okazakietal1987} as a generalised eigenvalue problem, subject to appropriate boundary conditions. (For more details on the numerical method and boundary conditions used see \citep{ferreiraogilvie2008}.) We find the simplest possible (lowest radial order, $l=0$) trapped mode to have an approximately Gaussian structure in $r$, centred at the maximum of $\kappa$, in accordance with \citep{perezetal1997} (see Fig. \ref{freer}).

\section{3. Excitation mechanism for trapped r modes}

We are interested in studying the growth of trapped inertial oscillations in deformed discs. Deformations like warping or eccentricity are typically identified with low-frequency modes with $m=1$, as these oscillations are global, long-lived, and vary on length-scales comparable to the characteristic disc radius. The warp is identified with an $(m,n)=(1,1)$ mode \citep{papaloizoulin1995}, while eccentric discs correspond to $m=1, n=0$ modes. Around rotating black holes, the stationary warp and eccentricity have an oscillatory radial structure \citep{ivanovillarionov1997} (see also Ogilvie \& Ferreira, this volume). The excitation mechanism for r modes relies on their interaction with these global deformations, which gives rise to an intermediate mode that can then couple with the warp or eccentricity to feed back on the original inertial oscillations.

The basic hydrodynamic equations have an intrinsic non-linearity through the term $\textbf{u}\cdot\nabla\textbf{u}$, which provides couplings between the different linear modes of the system. Here we focus on the couplings between the global deformations, and the inertial and intermediate modes that result in excitation of the simplest possible trapped r mode.

Before describing our numerical calculations on the excitation mechanism, we should analyse the energy exchanges between the different interacting modes, to make sure the r mode gains energy in this coupling. We consider the warp or eccentricity to have zero frequency so that it propagates everywhere in the disc (according to the dispersion relation), and therefore its energy is essentially zero. Consequently, the energy exchanges happen between the r and intermediate modes and the disc. Since the first is axisymmetric, it has positive energy, while the intermediate mode has negative energy, as it propagates inside its corotation radius. In the coupling between the global deformation and the r mode, the latter is amplified in the process of generating the negative energy intermediate wave. Sustained growth of the inertial mode requires this intermediate wave to dissipate, drawing positive energy from the disc rotation, so that its negative energy is continually replenished by the r mode. To achieve this dissipation, we choose to damp the intermediate wave locally at a rate $\beta\Omega$, where $\beta$ is a dimensionless parameter, i.e. we include a friction term in the equations for the intermediate mode. This does not necessarily mean that the mode is dissipated into the disc due to viscosity or friction. It is also a way of representing mathematically the fact that the intermediate mode might be absorbed at its corotation radius or at the marginally stable orbit.

For wave interaction to occur, the modes need to propagate in the same region of the disc, and the parameters $(\omega,m,n)$ need to obey certain coupling rules. For the interaction of the r mode $(\omega,0,1)$ with the warp $(0,1,1)$, two intermediate modes are possible: $(\omega,1,0)$ and $(\omega,1,2)$. The interaction with eccentricity $(0,1,0)$ results in only one intermediate mode $(\omega,1,1)$. For each one of the three possible interactions through which the inertial mode might be excited, we solve a system of equations for the r and intermediate modes, coupled by the warp or eccentricity \citep{ferreiraogilvie2008}. Even though the coupling terms arise from the non-linearities in the basic equations, the system is linear in the unknowns (the enthalpy and the components of the velocity of the r and intermediate modes). Expressions for warp or eccentricity enthalpy and velocity perturbations are determined beforehand \citep[see][]{ferreiraogilvie2008}.

Each of the systems of coupled equations is solved numerically as an eigenvalue problem, subject to certain boundary conditions. We aim to find solutions for which the radial velocity of the r mode resembles the function shown in Fig. \ref{freer}, i.e. we choose the coupling terms to be small enough that the eigenfunctions are not significantly altered, and the eigenvalues ($-\textrm{i}\omega$) change slightly by acquiring a small positive real part. The growth rate of the inertial mode is then given by the (positive) imaginary part of its frequency.

\begin{figure}
\centering
\begin{tabular}{cc}
\epsfig{file=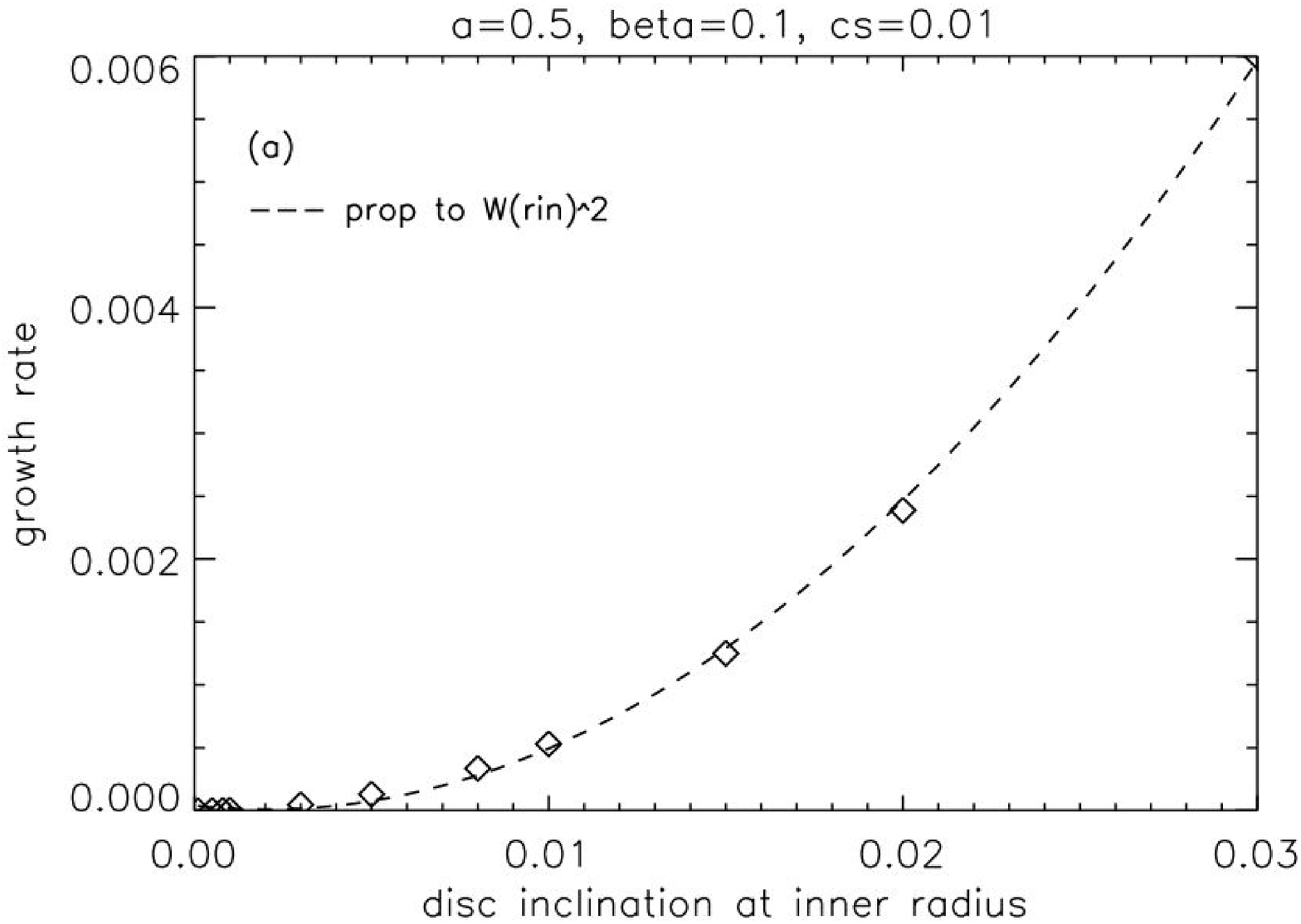,width=2.86 in,clip=} &
\epsfig{file=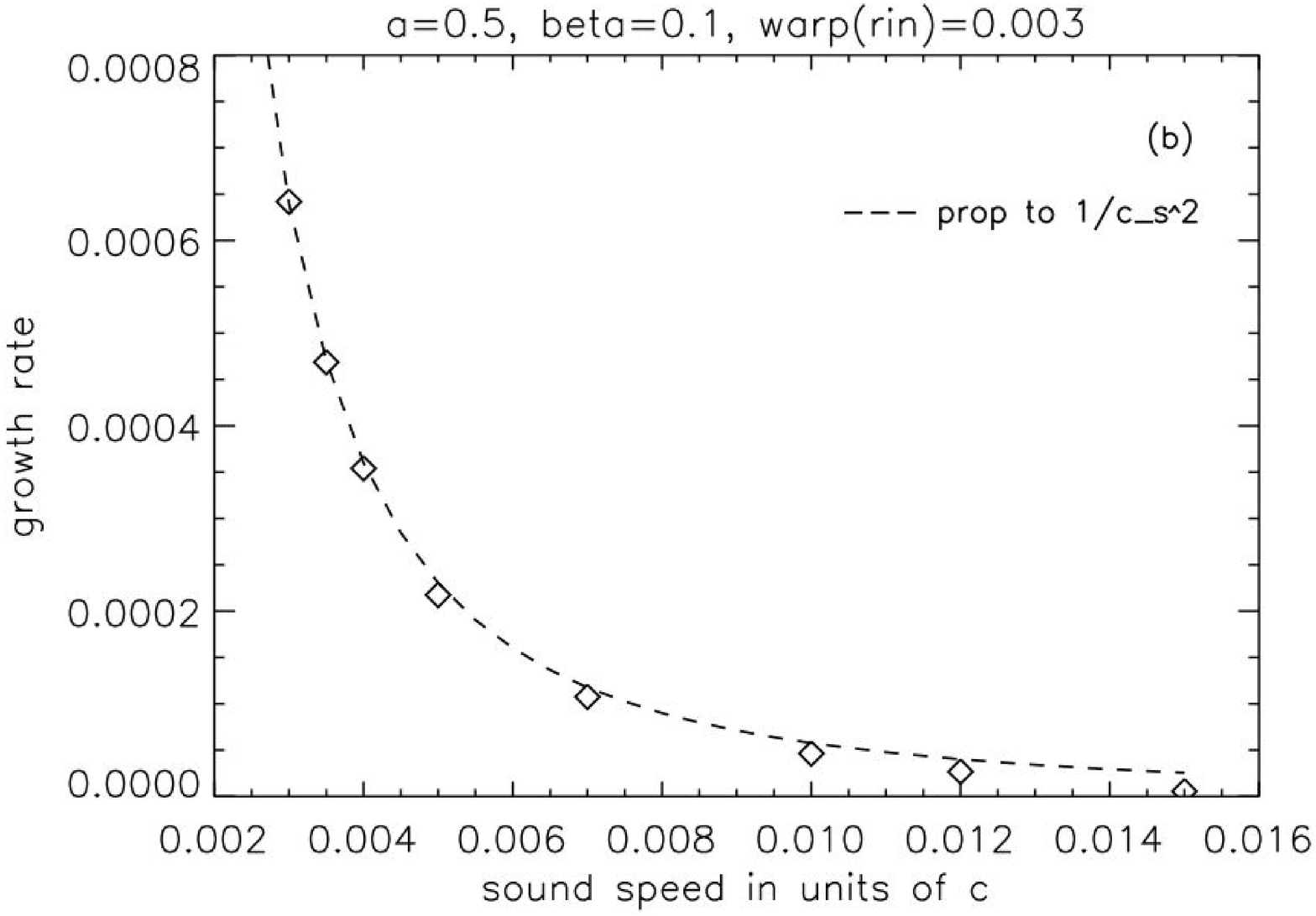,width=2.86 in,clip=} \\
\epsfig{file=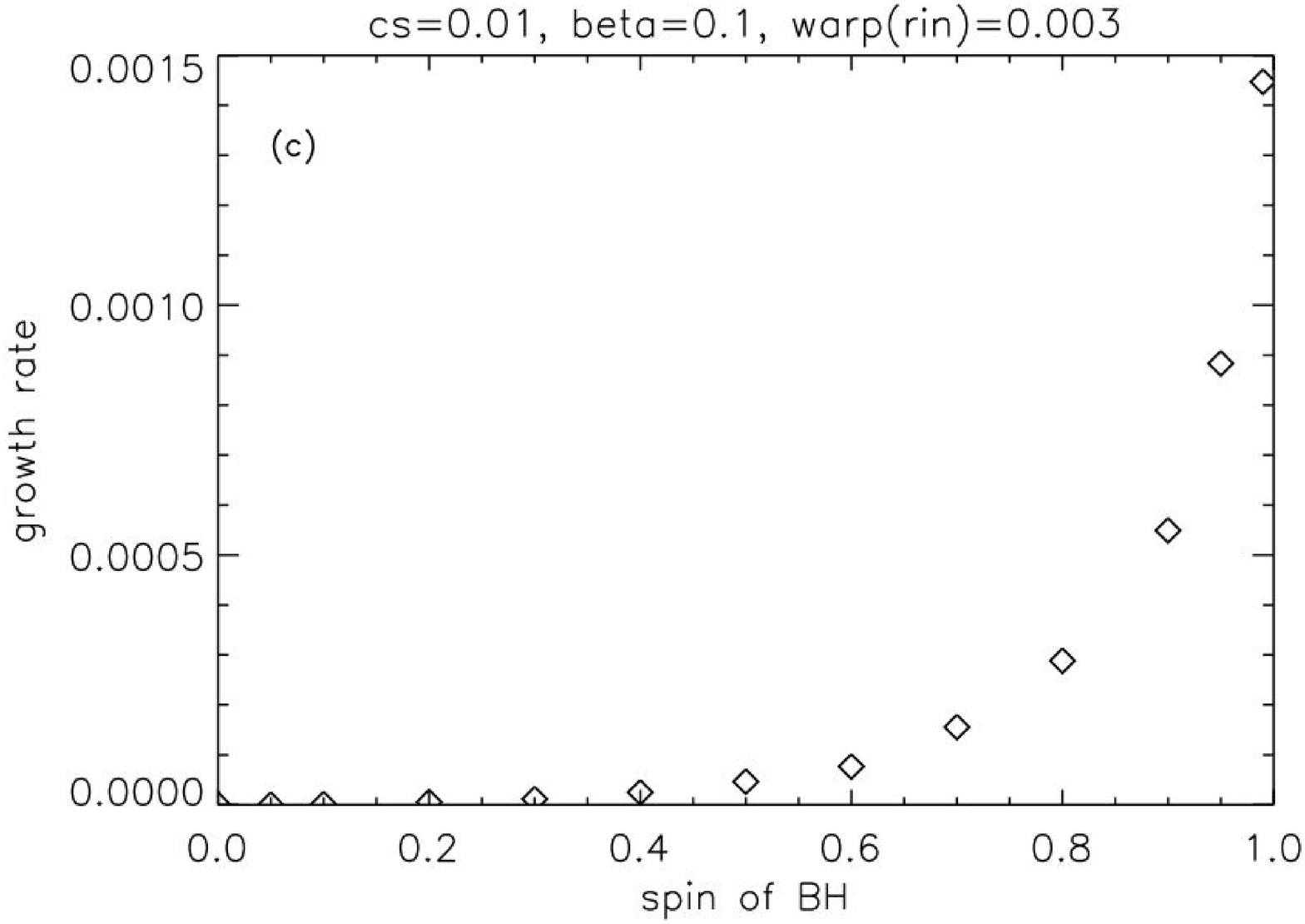,width=2.86 in,clip=} &
\epsfig{file=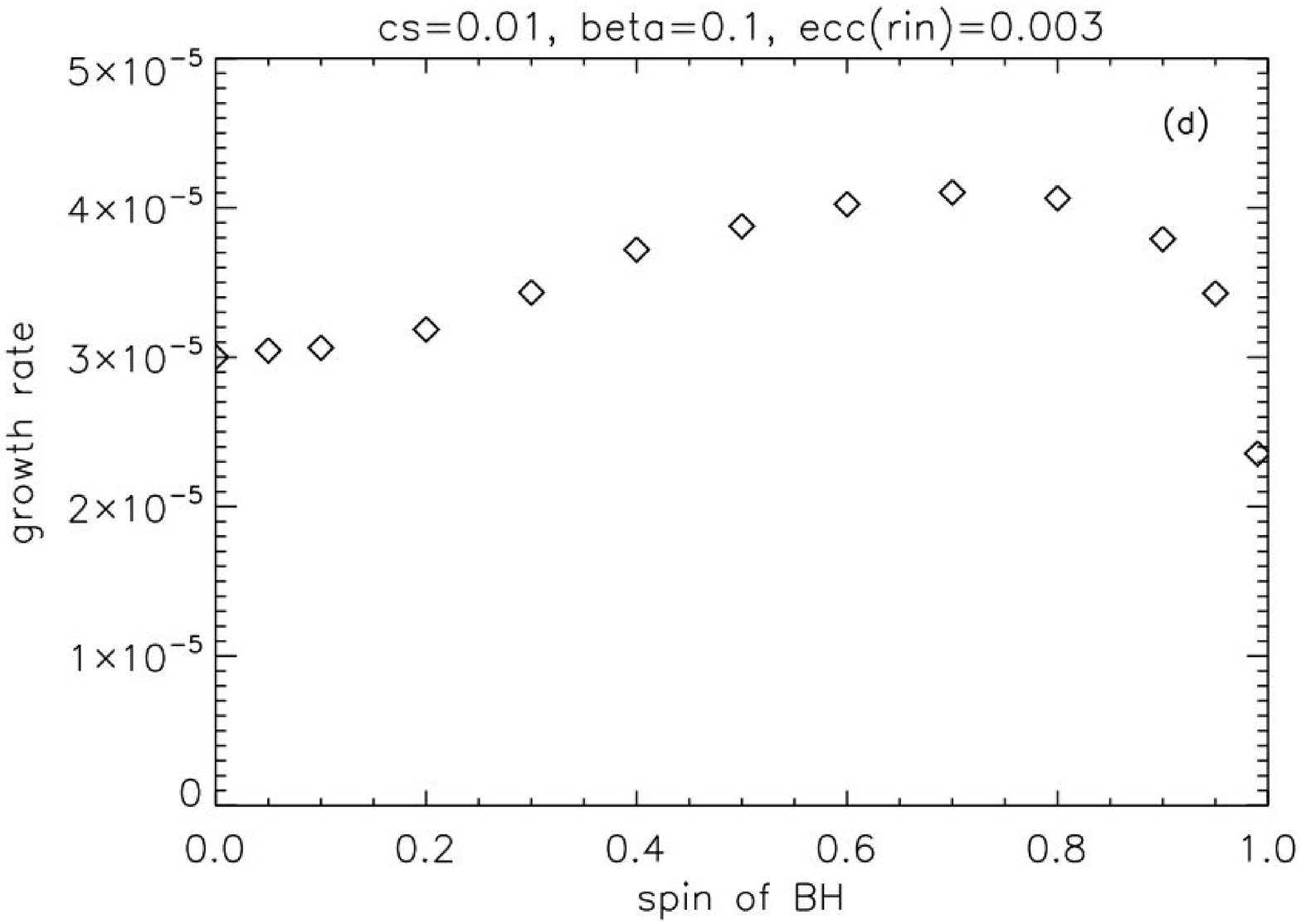,width=2.86 in,clip=} \\
\end{tabular}
\caption{\footnotesize{Panels (a) and (b) show the variation of the growth rate of the simplest trapped r mode, with warp amplitude at the inner radius, and sound speed in the disc, respectively, for the interaction with the $n=0$ intermediate mode. Panels (c) and (d) show the variation of the growth rate with the spin of the black hole, for the interaction with the $n=0$ intermediate mode in a warped disc, and the interaction with the $n=1$ intermediate mode in an eccentric disc, respectively.}}
\label{gr}
\end{figure}

In Figs \ref{gr} (a) and (b) we show the variation of the growth rate with the warp amplitude at the inner radius, and with the sound speed in the disc, respectively, for the interaction where the intermediate mode has $n=0$. Results are similar for the interaction with the $n=2$ mode, as well as for the interaction in the eccentric disc, in which case the warp amplitude at the inner radius is replaced by the eccentricity at the same location. In Figs \ref{gr} (c) and (d) we represent the variation of the growth rate with the spin of the black hole for the interaction of the r mode with the warp and $n=0$ mode (results for the $n=2$ mode are similar), and for the interaction in the eccentric disc. These results are discussed in the following section.

\section{4. Discussion}

In Fig. \ref{gr} (a) we can see that the growth rate of the inertial mode varies with the square of the disc inclination at the  inner radius. This is expected since the interaction relies on the use of the warp twice, and we can therefore argue that the growth rate is proportional to $|\textrm{d}W/\textrm{d}r|^2$, where $W(r)$ is the inclination of the disc at radius $r$ (and $\textrm{d}W/\textrm{d}r$ represents the actual warp). In turn, $|\textrm{d}W/\textrm{d}r|^2$ is proportional to the square of the inclination at the inner radius divided by the local wavelength of the warp ($\lambda_W$) squared, which, according to the dispersion relation for $(\omega,m,n)=(0,1,1)$, is proportional to $c_s^2$. Therefore, for a constant sound speed, we expect the growth rate to be proportional to the square of the inclination at the inner radius (Fig. \ref{gr} (a)). Also, if the latter is fixed, the growth rate is proportional to the inverse of the square of the sound speed, as we can see in Fig. \ref{gr} (b).

The variation of the growth rate with the spin of the black hole for the interaction in the warped disc (Fig. \ref{gr} (c)) shows that there is no mode excitation if the black hole is non-rotating. Although this result is in apparent contradiction with Kato's work \cite{katowarp2004}, we can argue that it is in fact expected. If the black hole is not rotating, the metric is spherically symmetric, and the warp is only a rigid tilt, which means that nothing changes in the disc except its inclination angle. If there are no relevant changes in the disc, there should be no relevant changes in the oscillations, i.e., a rigid tilt warp should not work as an excitation mechanism for waves in a disc. However, even in the case of a non-rotating black hole, it is still possible to have growth of oscillations, if the disc around it is eccentric (Fig. \ref{gr} (d)).

\section{5. Conclusion}

We have described an excitation mechanism for inertial modes trapped in the inner region of discs around black holes. The mechanism relies on a non-linear coupling between these waves and eccentricity or warping of the disc. Excitation of the inertial modes is possible if the intermediate mode resulting from the interaction deposits its negative energy into the disc, by being dissipated or absorbed at its corotation radius or the marginally stable orbit. Rotational kinetic energy is then removed from the disc and becomes available for the r mode to grow. Depending on the disc properties (sound speed and amplitude of the warp or eccentricity at the inner radius) and on the spin of the black hole, reasonable values of the growth rate can be achieved for a deformation of modest amplitude.


\begin{theacknowledgments}
We thank the organisers of `Cool discs, hot flows' for a very enjoyable and interesting conference. The work of BTF was supported by FCT (Portugal) through grant no. SFRH/BD/22251/2005.
\end{theacknowledgments}



\bibliographystyle{aipproc}   




\end{document}